\documentclass[a4paper,12pt,aps]{revtex4-1}
\usepackage{amsmath}
\usepackage{bbm}
\usepackage{graphicx,amssymb,bm,latexsym,color,epsf}
\usepackage{breqn}
\usepackage{cancel}
\usepackage[colorlinks]{hyperref} 

\newcommand{\be}{\begin{equation}}
\newcommand{\ee}{\end{equation}}
\newcommand{\bea}{\begin{eqnarray}}
\newcommand{\eea}{\end{eqnarray}}

\newcommand*{\eqVcenter}[1]{\vcenter{\hbox{#1}}}
\def\dt#1{{\buildrel{\hbox{\large.}}\over{#1}}}

\makeatletter
\let\cat@comma@active\@empty
\makeatother

\begin{document}
\title{On the renormalization of non-polynomial  field theories.}

\author{Andrea Santonocito}
\email{andreamariasantonocito@gmail.com}
\affiliation{Dipartimento di Fisica, Universit\`a di Catania,
via S. Sofia 64, I-95123, Catania, Italy;\\
Centro Siciliano di Fisica Nucleare e Struttura della Materia, Catania, 
Italy.}

\author{Dario Zappal\`a}
\email{dario.zappala@ct.infn.it}
\affiliation{INFN, Sezione di Catania, via S. Sofia 64, I-95123, 
Catania, Italy;\\
Centro Siciliano di Fisica Nucleare e Struttura della Materia, Catania, 
Italy.}

\begin{abstract}
\vskip 30pt
\centerline{ABSTRACT}
\vskip 10pt

A class of scalar models with non-polynomial interaction, 
which naturally admits an analytical resummation 
of the series of tadpole diagrams is studied in perturbation
theory. In particular, we focus on a model containing only one 
renormalizable coupling that appear as a multiplicative coefficient
of the squared field.
A renormalization group analysis of the  Green functions of the model
shows that these are only approximated solutions of  
the flow equations, with errors  proportional to powers of the
coupling, therefore smaller in the region of weak coupling.
The final  output of the perturbative analysis is that the 
renormalized model is non-interacting with finite mass and
vanishing vertices or, in an effective theory limited by an ultraviolet
cut-off, the vertices are suppressed by powers of the inverse cut-off. 
The relation with some non-polynomial interactions derived long ago, 
as solutions of the linearized functional renormalization group flow 
equations, is also discussed.

\end{abstract} 

\maketitle

\setcounter{page}{2}

\section{introduction}
\label{intro}

The Ultraviolet (UV) properties of the four-dimensional
scalar quantum field theories  are substantially under control
thanks to a  huge amount of results and indications coming not only 
from the standard perturbative techniques \cite{CJP,IIM}, 
but also from different non-perturbative approaches
including formal investigations\cite{AIZ,FRO}, 
or Montecarlo simulations of lattice  
theories\cite{LW1,LW2,LW3,HJJL}, 
or Renormalization Group (RG) analysis\cite{Wilson:1973,Hasenfratz,zinn,
MorrisReb,Morris:1996,Morris:1998}, 
all pointing toward the existence of a single fixed point (FP), 
the  Gaussian FP, which admits only a renormalized free field theory,
or equivalently an  effective scalar field theory whose validity 
is limited by an UV cut-off,
that shows an interaction strength which grows at larger energy scales 
up to a Landau-like pole. 

Despite this tight scenario, the activity of searching for
alternative mechanisms, capable of reshaping the UV structure 
associated to scalar interactions, has been (and still is) 
quite active, not only to  refine or establish novel 
non-perturbative techniques on a more formal side,
but also to investigate on any possible modification of the 
high energy picture of  all contexts where scalar fields are involved,
from the UV completion of the Higgs field  in the 
Standard Model to the effective description of the inflaton field
in cosmology \cite{
HHOrig,Periwal:1995,HHReply,HH2,Bonanno:2000,Branchina:2000,
Gies:2000,Alt:2004,Alt:2005,Gies:2009,Pietr:2012,Holdom:2014,Gies:2015,
Gies:2016,Morris:2016,Morris:2022}.

An old attempt 
to renormalize quantum field theory by enlarging 
the derivative sector of the action through the introduction of 
additional terms containing  higher space-time derivatives of the 
field, \cite{thirring,Pais,Stelle}, was reconsidered more recently 
both for scalar models that present space and time  derivatives  
in equal number, and for anisotropic 
models where the number of space derivatives in the action is larger 
than the time derivatives. 
This is due to the basic principle 
that an increase of the number of derivatives raises the power
the momentum in the propagator, thus reducing the global degree 
of divergence of a generic quantum correction diagram.

The  renormalization of these models, represented 
in Euclidean space, can be traced back to the structure 
of the associated Lifshitz FPs,  that may appear if higher 
derivative terms are present in the action \cite{Horn,hornrev,Diehl3}.
So, for instance, the isotropic case with four space and four time 
derivatives in four dimensions presents a non-trivial  phase 
structure, with a line of Lifshitz FPs, that shares many properties 
with the two-dimensional Kosterlitz-Thouless transition
\cite{Bonanno:2014,Zappala:2017,
Zappala:2018,Zappala:2019,Defenu,Buccio}. 

However, since  the presence of time derivatives of order larger 
than two leads to the Ostrogradski instability, 
associated with Hamiltonians unbounded from below 
which violate unitarity\cite{deUrries:1998}, 
it is preferable to maintain only two time derivatives thus avoiding
any problem with unitarity in the UV sector. 
This point of view leads to the anisotropic case where only two 
derivatives are kept in one  direction
(Euclidean time) and a larger number of derivatives is reserved to the 
other (space) directions. This scheme was proposed in 
\cite{horava} to make the gravitational action renormalizable, but it 
is also applied to the study of the UV sector of various field 
theories \cite{Anselmi:2007,Iengo,
Dhar:2009,horava:ym,alexandre,Chao:2009,Solomon:2017nlh,zap1,zap2}.
In fact, even in anisotropic form,
a sufficiently high number of space 
derivatives improves the UV behavior of the theory to the point of
converting a UV unstable Lifshitz point into a stable one.

In the anisotropic  framework,
the role of tadpole diagrams is crucial as, for a 
sufficiently large number of space derivatives,  
only this class of diagrams is UV divergent, which makes the 
renormalization procedure easier to implement. Moreover, as noticed in
\cite{alexandre,zap1,zap2}, under specific assumptions on the 
couplings of the 
theory, it is possible to sum the whole series of tadpole diagrams
into a compact form and also to prove that the corresponding theory 
is asymptotically free. 
Clearly,  this  renormalization property
of the tadpole diagrams follows from the specific structure of the 
derivative sector, but it is worthwhile to investigate on the 
the analogous mechanism for a scalar action with standard
derivative sector (i.e. two space and two time derivatives), 
especially because the summable tadpole diagram series comes from 
a potential that is non-polynomial in the fields and, in principle,
one expects these interactions to be non-renormalizable.

Therefore, in this paper we analyse a scalar toy model with 
standard derivative sector and  non-polynomial interaction, 
that is representative of a class of models which  naturally 
lead to a summable series of tadpole diagrams, to find 
out whether the mentioned property could lead to interesting 
consequences even in this case. It  turns out that 
the sum of the tadpole  series is sufficient to guarantee
the perturbative renormalizability of such a  model, 
although generating a renormalized free scalar theory.

In addition, it must be recalled that a study of the UV sector of
scalar  non-polynomial theories in four dimensions was 
conducted long ago  in a different context, 
namely the Renormalization Group (RG) analysis of the 
Wilsonian action, and the conclusion was achieved
that the  differential flow equation, suitably linearized around
the Gaussian FP, admits a class of relevant solutions,
(i.e. solutions that fall into the the Gaussian FP, when the 
UV limit of the RG energy scale $k\to\infty$ is taken), in 
contrast with the well settled picture of the trivial 
scalar theory \cite{HHOrig,HHReply,HH2}.  In addition, these solutions 
can be expressed in the form of a non-polynomial expansion in 
powers of the field and, they show many similarities with the 
toy model here considered.

After some debating 
\cite{MorrisReb,Morris:1996,Morris:1998,
HHOrig,Periwal:1995,HHReply,HH2,Bonanno:2000,Branchina:2000,
Gies:2000,Alt:2004,Alt:2005,Gies:2009,Pietr:2012,Morris:2016}, 
these solutions were considered incorrect because 
(see \cite{Morris:2016} for a definite explanation),
the assumption of uniform smallness of the solution, 
which is essential when the linear  version of the RG flow 
equation is considered, is in fact violated, at least at 
large  values of the field and, consequently, their flow is not 
correctly predicted by the linear RG flow equation and the 
conclusion that they represent asymptotically free interaction
is wrong. Then, a comparison of this RG solution
with the toy model here considered is mandatory and we 
need to analyse the RG flow of the latter and point out 
the differences with the former.

After introducing the scalar toy model and studying its
main properties in perturbation theory in Sec. \ref{sec1} (and 
those of similar models in Appendix \ref{appB}), 
in Sec. \ref{sec2} we pass to a description of the Halpern-Huang 
solutions and to an investigation on their limits through a 
diagrammatic analysis. Then, Sec. \ref{sec3} we apply the RG 
machinery to our scalar model and show the level of 
approximation at which it can be taken as a solution 
of the full flow equation. In addition, the relation 
with the  solution of \cite{HHOrig,HH2} is discussed.
Conclusions are reported in Sec. \ref{Conclusions}.

\section{non-polynomial toy model in perturbation theory}
\label{sec1}

\subsection{Main characteristics of the model}
In this section we focus on the particular model, whose Euclidean action in four 
dimensions, 
$ S_E=\int {\rm d^4 } x \;\; \left ( \frac{1}{2} \partial_\mu \phi \;
\partial_\mu \phi  + V (\phi) \right ) $
has the following non-polynomial potential 
\begin{equation}
 V(\phi) = M^4 \left( \frac{g_0 \phi^2}{2! M^2} + \frac{g_0^2 \phi^4}
 {4! M^4} + \frac{g_0^3 \phi^6}{6! M^6} + \dots \right) = M^4 \left[ 
 \cosh \left(\sqrt{g_0} \frac{\phi}{M}\right) - 1 \right]
 \label{potential}
\end{equation}
where $M$ is a fixed (not subjected to renormalization) 
mass scale, $g_0>0$ is the bare coupling constant and the  field 
independent term is set to zero. The structure of $V(\phi)$ is non-
polynomial: it is a series where any even power of the field $\phi$ is 
included and the coefficients are arranged in such a way that the 
sum of the series is a known function, namely the hyperbolic cosine
of $\left (\sqrt{g_0}\, \phi /M \right )$.
This implies that there is only one independent parameter $g_0$
(besides $M$), but the powers of $g_0$ increase proportionally to the 
powers of the field $\phi$ in the various terms of the potential,
and this  allows us to arrange the diagrams in a perturbative series 
in powers of $g_0$. Another peculiar feature is that the only 
dimensional parameter of Eq. (\ref{potential}) is $M^2$ and, 
in particular, the bare square mass of the theory is $g_0 M^2$; 
as a consequence the dimensional content of any 
renormalized quantity can be  expressed in terms of $M$. 

If we want to compute the quantum corrections of the Green Functions, 
first we need to determine the degree of divergence of each
diagram which is different from the standard results, as in this case
we deal with  vertices with any possible (even) number of legs. 
Therefore we  introduce the notation:
$P =$ total power of $g_0$ of a diagram;
 $N =$ number of legs of a specific vertex;
$E =$ number of external lines of a diagram;
$H =$ number of internal lines of a diagram;
 $L =$ number of loops of a diagram;
$V_N =$ number of vertices with $N$ legs of a diagram;
$V = \sum_{\text{vertices}} V_N = $ total number of vertices
of a diagram.
The following relations hold:
\begin{equation}
	P = \sum_{\text{vertices}} \frac{N}{2} \qquad \qquad 
 \sum_{\text{vertices}} N = E + 2H \qquad \qquad L = H - V + 1
 \label{degreeod}
\end{equation}
and we notice  that the total power $P$ of each diagram is not just
the total number of vertices, but it depends on the specific vertices 
that enter the diagram itself. 
By combining  the first two equations, we get
$	P = {E}/{2} + H$ and finally the superficial degree of divergence 
of a diagram, measured as the resulting power of the momentum cut-off 
$\Lambda$
used to regulate the UV divergent integrals, is given by:
\begin{equation}
\label{sdd}
\mathcal{D}_\Lambda = 2P - E - 4V + 4
\end{equation}
in contrast to the  standard quartic interaction result, 
$\mathcal{D}_\Lambda = 4- E$. 
Eq. (\ref{degreeod}) indicates that, for a given $E$-point Green 
function, at a fixed perturbative order in $g_0$, i.e. at fixed $P$,
the most divergent contribution comes from the diagrams with minimum 
$V$, which clearly correspond to the tadpoles with $V=1$, and we 
notice that, due to the presence of the tower of couplings appearing 
in Eq. (\ref{potential}), diagrams with $V=1$ could still contain an 
arbitrary number of tadpoles.

This property can now be exploited when considering the order by order 
renormalization of the model. We start from the 2-point  function 
which, up to the fourth order in $g_0$, reads:
\begin{equation}
\label{twopgf}
	\begin{gathered}
		\eqVcenter{\includegraphics{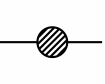}} = 
  \underbrace{\eqVcenter{\includegraphics{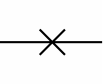}}}_{g_0} + 
  \underbrace{\eqVcenter{\includegraphics{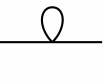}}}_{g_0^2} + 
  \underbrace{\eqVcenter{\includegraphics{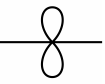}}}_{g_0^3} + \\
		+ \underbrace{\eqVcenter{\includegraphics{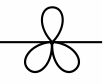}} + \eqVcenter{\includegraphics{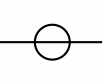}}}_{g_0^4} + \, \dots = \\
= g_0 M^2 + g_0^2 \, I + g_0^3 \, \frac{1}{2!} \, \frac{I^2}{M^2} + 
g_0^4 \left( \frac{1}{3!} \, \frac{I^3}{M^4} + S \right) + O(g_0^5)
\end{gathered}
\end{equation}
In Eq. (\ref{twopgf}) the cross
indicates the zero-th order vertex and $I$ denotes half of the tadpole 
integral, 
which can be computed by means of a four-momentum cut-off $\Lambda$
\begin{equation}
\label{eq:I}
I \equiv \eqVcenter{\includegraphics{2p_1t}} = \frac{1}{2} \int \frac{d^4k}
{(2 \pi)^4} \, \frac{1}{k^2+m^2}=
 c \left( \Lambda^2 - m^2 \log \frac{\Lambda^2}{m^2} \right)
\end{equation}
where $c=1/(32 \pi^2)$ and $m$ indicates the renormalized mass, 
while $S(p)$ is the sunset integral  with external momentum $p$:
\begin{equation}
\label{eq:S}
	S(p) \equiv \eqVcenter{\includegraphics{2p_2m}} = \frac{1}{3!} \int 
 \frac{d^4a}{(2 \pi)^4} \int \frac{d^4b}{(2 \pi)^4} \, \frac{1}{a^2+
 m^2} \, \frac{1}{b^2+m^2} \, \frac{1}{(a+b+p)^2+m^2}
\end{equation}
that, with the same regularization, can be written as (see Eq. \eqref{sunset} in Appendix \ref{appA})
\begin{equation}
\label{eq:SS}
	S(p)= \frac{1}{6 (4 \pi)^2} \left[ 3 \Lambda^2 + p^2 \left( \alpha
 + \beta \log \frac{\Lambda^2}{m^2} \right) + O \left(\log 
 \frac{\Lambda^2}{m^2}\right) + O(p^2,m^2) \right]
\end{equation}
with $\alpha$ and $\beta$ constants. It must be noticed that the 
most divergent contribution of the tadpole integral is 
$I_\Lambda \sim \Lambda^2$, as for  the sunset integral. 
Thus, in the fourth order diagrams shown in Eq. (\ref{twopgf}), 
the diagram with 3 tadpoles is far more divergent that the sunset and 
this is just a particular case of the more general property discussed 
above that,  at any given order in perturbation theory, the most 
divergent diagram is the one with one vertex, namely the one 
consisting of a product of tadpoles.

Then, in a perturbative scheme, we want to write $g_0$ as a series 
expansion of the renormalized coupling $g_R$:
\begin{equation}
	\label{g0_renseries}
	g_0 = g_R + g_R^2 \, \delta_1 + \frac{1}{2!} g_R^3 \, \delta_2 + 
 \frac{1}{3!} \, g_R^4 \, \delta_3 + O(g_R^5)
\end{equation}
where $\delta_1$, $\delta_2$, $\delta_3$, \dots are counterterms
that must be fixed so that the Green Functions remain finite in the 
limit $\Lambda\to \infty$. At order $O(g_0)$ we define:
\begin{equation}
\label{delta1}
	\delta_{1} = \frac{c}{M^2} \left( \mu^2 - \Lambda^2 - m^2 
 \log \frac{\mu^2}{\Lambda^2} \right)
\end{equation}
where $\mu$ is a finite arbitrary scale to be fixed by a 
renormalization  condition, so that the dangerous terms when 
$\Lambda\to \infty$  get cancelled:
\begin{equation}
\label{rencond}
J(\mu,m^2) =  I + \delta_1 M^2 = 
c \left( \mu^2 - m^2 \log \frac{\mu^2}{m^2} \right) 
\end{equation}

Moreover, at higher order, we choose:
\begin{equation}
\label{delta23}
\delta_2 = \delta_1^2 - 2 \, \delta_1 \, \frac{I}{M^2} \qquad \qquad 
\delta_3 = \delta_1^3 + 6 \, \delta_1 \, \frac{I^2}{M^4} - 9 \, 
\delta_1^2 \, \frac{I}{M^2} - 6 \, \frac{S(p^2=\mu^2)}{M^2}
\end{equation}
where the external momentum $p$ in  Eq.(\ref{eq:SS}) is set equal 
to the renormalization scale $\mu$ and 
the expansion in Eq. (\ref{twopgf}) takes 
the simple form:
\begin{equation}
	\label{eq:2p_rc}
\eqVcenter{\includegraphics{2p_complete}} = g_R M^2 + g_R^2 \, J + g_R^3 \, 
\frac{1}{2!} \, \frac{J^2}{M^2} + g_R^4 \, \frac{1}{3!} \, \frac{J^3}
{M^4} + O(g_R^5) \;.
\end{equation}
Then, any divergent term in the limit $\Lambda\to \infty$ is 
cancelled. With the help of Eq. (\ref{delta23}), 
and by retaining only the most divergent contributions of the integrals 
$I$ and $S$, the expansion in Eq. (\ref{g0_renseries}) becomes
\begin{equation}
	g_0 \sim g_R - g_R^2 \, \frac{\Lambda^2}{M^2} + g_R^3 \, \frac{3}
 {2} \, \frac{\Lambda^4}{M^4} - g_R^4 \left( \frac{8}{3} \, 
 \frac{\Lambda^6}{M^6} - 6 \, \frac{\Lambda^2}{M^2} \right)
\end{equation}
and the second term in brackets, coming  from the sunset diagram,  
can be neglected. 

Furthermore, if we discard, order by order,
all terms with non-leading powers of $\Lambda$, we end up with the 
sum of the multiple tadpole diagrams which, as already discussed, 
provide at  each order, the contribution with the largest power 
of $\Lambda$. This is an essential point. In fact, the various 
numerical factors of the couplings of the potential in 
(\ref{potential}) are suitably chosen so that the tadpole series can 
always be summed to an exponential function as for instance in the case 
of the six-point vertex:
\begin{equation}
\label{sumtadpole}
	\eqVcenter{\includegraphics{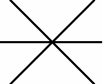}} + \eqVcenter{\includegraphics{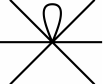}} +
 \eqVcenter{\includegraphics{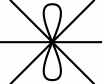}} + \; \dots \; = \frac{g_0^3}{M^2} \, 
 e^{g_0 I / M^2} \; .
\end{equation}
The same holds true for any $2n$-point vertex. Moreover, it 
is simple to 
show by induction that any vertex of a generic diagram can be dressed 
by the exponential tadpole series as indicated below, for instance,
for one vertex of the sunset diagram
\begin{equation}
	\eqVcenter{\includegraphics{2p_2m}} + \eqVcenter{\includegraphics{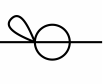}} + 
 \eqVcenter{\includegraphics{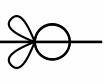}} + \; 
 \dots \; = g_0^4 \, e^{g_0 I / M^2} S
\end{equation}

Therefore, we shall exploit this property to set a renormalization 
condition for the coupling $g$, and we shall check that this is 
sufficient to fully renormalize the model. The renormalization 
condition is directly read from the tadpole dressing of the mass term
\begin{equation}
\label{2pttad}
	\eqVcenter{\includegraphics{2p_0t_crossed}} + \eqVcenter{\includegraphics{2p_1t}} 
 + \; 
 \dots \; = M^2 g_0 \, e^{g_0 I / M^2} = M^2 g_R \, e^{g_R J / M^2}
\end{equation}
i.e. from the following equation that involves both $g_0$ and the 
renormalized coupling $g_R$:
\begin{equation}
	\label{eq:renormalization}
	g_0 \, e^{g_0 I / M^2} = g_R \, e^{g_R J / M^2}
\end{equation}
It is understood that the right hand side of 
\eqref{eq:renormalization} does not depend on the cut-off
$\Lambda$ and since, according to \eqref{rencond},
$J=J(\mu,m^2)$,  it follows $g_R=g_R(\mu)$,
which guarantees
the $\mu$-independence of the left hand side of 
\eqref{eq:renormalization}.

Eq. (\ref{eq:renormalization}) has the structure of the Lambert 
equation $z = w e^w$, with  complex $w$ and $z$, that admits a 
multi-valued solution, made by the branches of the the Lambert 
function $w_k(z)$ (where $k$ indicates the specific branch). In the 
case of real $w$ and $z$, the principal branch of the Lambert 
function $w_0(z)$ exists only in the range $z\geq 0$ and 
$-1/e\leq z< 0$ , and the large $z\to \infty$ behavior of $w_0(z)$ is 
\begin{equation}
\label{eq:LF}
w_0(z) \sim \log z - \log \log z 
\end{equation}

If we treat the right hand side of Eq. (\ref{eq:renormalization})
as a fixed quantity computed  at a particular value of $\mu$,
and identify $({I}/{M^2}) \, g_0 $  with the Lambert function $w_0$
and also identify $(g_R\, I/M^2)\; {\rm exp}\,[ g_R \,J / M^2)] $ with 
the variable $z$, we  derive the dependence of $g_0$ on the cutoff 
$\Lambda$ directly  from (\ref{eq:LF}). In fact, by recalling that 
$I \sim \Lambda^2$, and  defining
$\hat{\Lambda}^2 = {\Lambda^2}/{M^2}$,
we find, for $\hat{\Lambda}^2 \to \infty$,
\begin{equation}
	g_0 \sim \frac{\log \hat{\Lambda}^2 - \log \log \hat{\Lambda}^2}
 {\hat{\Lambda}^2}
 \label{g_0asy}
\end{equation}
and it is easy to check that Eq. (\ref{g_0asy}) is 
consistent with our renormalization condition 
(\ref{eq:renormalization}) 
\begin{equation}
\label{finit}
	g_0 \, e^{g_0 \,  I / M^2} = \text{finite} 
\end{equation}
Eq. \eqref{finit} in turn  implies:  
${1}/{g_0} \sim {\rm exp}\,[g_0\, I / M^2]$, when $\hat \Lambda\to \infty$.

It is worth noticing that also the  mass renormalization condition, 
obtained from the $2$-point Green Function  by taking the 
renormalizations scale $\mu=m$,
\begin{equation}
	 m^2 = M^2\; g_R(m^2) \, e^{g_R(m^2) \, J(m,m^2) / M^2}    
\end{equation}
reduces to a Lambert equation for $g_R(m^2)$, once the  two 
scales $m$ and $M$ are chosen.

\subsection{Renormalization to all orders}
The next step consists in showing that the renormalization 
prescription in (\ref{eq:renormalization}) ensures the 
cancellation of all divergent terms. To this purpose we reconsider 
Eq. (\ref{sdd}) where the superficial degree of divergence of a generic 
diagram is displayed and check how it gets modified when we dress 
each vertex of the diagram by the full series of tadpoles 
and, consequently, include the cut-off dependence of $g_0$ 
given in Eq. (\ref{g_0asy}).

In other words, we recalculate the degree of divergence of 
a generic diagram, after inserting the full tadpole series 
at each vertex of the original diagram and after renormalizing each 
coupling $g_0$ according to Eqs. (\ref{eq:renormalization}) and 
(\ref{g_0asy}). Then, the diagram shows  the 
following cut-off dependence for $\hat \Lambda \to \infty$ 
\begin{equation}
	\label{eq:trend_generic_diagram}
	\left( \frac{\log \hat{\Lambda}^2 - \log \log \hat{\Lambda}^2}
 {\hat{\Lambda}^2} \right)^{P-V} \hat{\Lambda}^{2P-E-4V+4} = 
 \frac{\left(\log \hat{\Lambda}^2 - \log \log \hat{\Lambda}^2\right)^{P-
 V}}{\hat{\Lambda}^{E+2V-4}}
\end{equation}
where the term in brackets in the left hand side comes from the $P$ 
contributions corresponding to the total count of couplings $g_0$ 
of the diagram, minus the number $V$ of vertices, because the 
tadpole dressing at each vertex transforms one factor $g_0$
into a $\hat \Lambda$-independent, renormalized coupling.

If we neglect the logarithmic corrections in Eq. 
(\ref{eq:trend_generic_diagram}), for any diagram with fixed $E$ 
and $V$,  the trend for large $\hat{\Lambda}$ is the same, 
regardless of the form of the specific vertices that enter the 
diagram itself.  This is because any increment by one 
power of $g_0/M^2$ in  any vertex  requires
 to be compensated, at dimensional level, by a factor $\Lambda^2$; 
however, as $g_0 \sim  1/\hat{\Lambda}^2$, in the end there is no 
change in the overall  trend of the diagram. 
Logarithmic divergences not taken into account so far, will be 
analyzed  below.

From Eq. (\ref{eq:trend_generic_diagram}), it is clear that 
a larger number of vertices, as well as a larger number of
external legs, favours the convergence of the diagram
and, as we have already covered (and renormalized) the 
one-vertex diagrams, the worst possible scenario is represented 
by the case ($V=2$,  $E=2$). For this type of diagrams, 
by neglecting  the logarithmic contribution, the cut-off 
dependence is proportional to $1/\hat{\Lambda}^2$, thus they give zero 
contribution when the limit $\hat{\Lambda} \to \infty$ is taken. 
Green functions with larger $E$ and diagrams with $V\geq 2$ 
have more effective suppressing factor. 
As  a further  example, the $O(g_0^9)$ diagram
\begin{equation*}
	\eqVcenter{\includegraphics[scale=1.1]{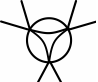}}
\end{equation*}
has $\mathcal{D}_\Lambda = 4$ and, after  dressing   each vertex
with the tadpole series and after renormalizing the coupling $g_0$,
we are left with a suppressing factor $g_0^6\sim 1/\hat\Lambda^{12}$.
Again, logarithmic corrections have not been taken into account.

However, because of the characteristic $\hat\Lambda$ dependence 
displayed in Eq. (\ref{eq:trend_generic_diagram}),
the logarithmic corrections within a specific series of diagrams
could in principle sum up to a power-like divergent factor,
which could potentially lead to an overall divergence.
In order to show that this scenario is to be excluded,
we concentrate again on the worst possible case.
It is easy to realize that the set of diagrams where this effect 
is maximized, corresponds to the first column in the grid shown below
\begin{equation}
	\begin{matrix}
		\eqVcenter{\includegraphics{2p_2m}} \qquad & 
  \eqVcenter{\includegraphics{2p_2m_1t}} \qquad & 
  \eqVcenter{\includegraphics{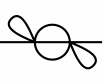}} \qquad & \dots \\
		\eqVcenter{\includegraphics{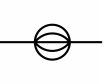}} \qquad & 
  \eqVcenter{\includegraphics{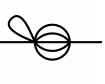}} \qquad & 
  \eqVcenter{\includegraphics{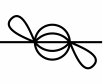}} \qquad & \dots \\
  \eqVcenter{\includegraphics{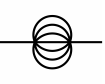}} \qquad & \eqVcenter{\includegraphics{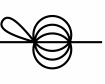}} 
  \qquad & \eqVcenter{\includegraphics{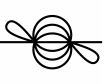}} \qquad & \dots \\
   \dots \qquad & \dots \qquad & \dots \qquad & \dots \\
	\end{matrix}
 \label{matrix}
\end{equation}
In fact, along the first column, the sunset diagram on the 
top-left corner
is dressed with the progressive insertion of one internal line 
that connects the two vertices, thus providing a melonic-like series of 
diagrams, all with ($V=2$, $E=2$) which, according to 
Eq. (\ref{eq:trend_generic_diagram}),
gives the largest contribution (apart from the one-vertex diagrams)
in the UV limit  $\hat\Lambda\to \infty$. Then, along each row,
at both vertices of the diagram on the left side of the row, 
we consider the dressing with the  tadpole series 
whose divergent sum is made finite by one power of the coupling $g_0$,
as already shown.

To prove that the full sum of the diagrams in (\ref{matrix}) 
does not produce any dangerous
divergence generated by the sum of the logarithmic terms shown in 
Eq. (\ref{eq:trend_generic_diagram}), we proceed by first  summing and 
renormalizing the tadpole series along the rows of  (\ref{matrix}),
and then by making  use of the following result, derived  in Appendix \ref{appA} in Eq. \eqref{inequality} ($b$ is a numerical constant),
\begin{equation}
	\begin{gathered}
\hat{S} = \eqVcenter{\includegraphics{2p_2m}} + \eqVcenter{\includegraphics{2p_4m}} + 
\eqVcenter{\includegraphics{2p_6m}} + \dots \le \\
		g_0^2 I + \frac{b M^4}{I} \left[ \cosh 
  \left( g_0 \frac{I}{M^2} \right) - 1 - \frac{1}{2} g_0^2 \frac{I^2}{M^4}
  \right]
	\end{gathered}
 \label{maggior}
\end{equation}
i.e. the melonic-like diagram series (where it is understood 
that the full tadpole series is included to dress each vertex), 
turns out to be smaller than the right hand side in (\ref{maggior}).

It is then straightforward to check that each term of the
right hand side in (\ref{maggior}) vanishes in the limit 
$\hat\Lambda \to \infty$, according to Eqs. (\ref{eq:I}) and 
(\ref{g_0asy}) (note that the ${\rm log \, log} \hat \Lambda^2$
in (\ref{g_0asy}) is essential to compute the correct  limit of the 
hyperbolic cosine).
Then, at least in this case, no divergent term is generated by the sum 
of the  logarithmic terms.

Actually, the same series of melonic-like diagrams shows up  when computing 
the $4,6,8,10,\dots$-point Green functions, even though the series in those 
cases have different numerical coefficients (generally higher) in front, 
because of the possible permutations of the external legs. 
In addition, all possible insertion of different vertices must be counted: 
for example the 6-point Green function can be constructed with 
$3$ external legs at each of the two vertices, but also with  $1$ leg in 
one vertex and $5$ legs in the other, and so on. Nevertheless, for these 
diagrams the suppressing factor is even stronger, because the power of $g_0$ 
is higher and  because a larger $E$ must be inserted in Eq.  
(\ref{eq:trend_generic_diagram})). Therefore, all contributions 
to the $2n$-point Green functions from the melonic-like series in 
(\ref{matrix}) vanish.

Moreover, concerning the 
$2n$-point Green functions with $n>1$,
there is the contribution of a similar series, 
see for instance the 4-point Green function  
\begin{equation}
\label{fish}
	\eqVcenter{\includegraphics{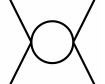}} + \eqVcenter{\includegraphics{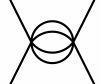}} + 
 \eqVcenter{\includegraphics{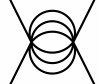}} + \; \dots
\end{equation}
But, in this case the first diagram of the series 
is only logarithmically divergent (in contrast to the sunset diagram 
which 
is quadratically divergent). So, each diagram of the series is less 
divergent than the corresponding in (\ref{maggior}) and consequently, 
as in the latter case, we find vanishing contribution.
Then, even  in the worst-case scenario (namely the 2-point melonic 
resummation) the logarithmic contributions sum up to a function that is 
still less 
divergent than the suppressing powers of $1/\hat{\Lambda}$ coming from 
the dressed 
vertices; all other cases are even more suppressed. 

Therefore, only the 2-point  Green function of this model is finite,  due to
the tadpole series, while all other Green functions  are vanishing:  we 
conclude that the theory is non-interacting, yet it gets a mass 
renormalization.

Finally, in Appendix \ref{appB} we discuss some models constructed 
by introducing some modifications with respect to the one  in 
\eqref{potential}, but in all cases considered
we end up either  with non-renormalizable theories or with structures 
that are equivalent to the one studied here.


\section{RG flow in Linear approximation}
\label{sec2}

Now, we reconsider the solutions of the linearized RG 
flow equation first pointed out in \cite{HHOrig,HHReply,HH2}, 
as  they share some crucial properties with 
the model discussed in Sect. \ref{sec1}. For our purposes, it is 
sufficient to employ  the Wegner-Houghton equation that describes 
the flow of the running potential $U_k(\phi)$ of a Wilsonian action, in 
terms of a coarse  graining scale $k$. In fact, 
as discussed in \cite{Morris:2016},
the linearized equation derived from the Polchinski flow or
from the Legendre effective average action flow in the Local Potential 
approximation, turn our to be equivalent to the previous one.

So, the Wegner-Houghton equation reads:
\begin{equation}
\label{wh}
	k \partial_k {U}_k(\phi) = - 2 c k^4 \log \left( 1 + 
    \frac{U''_k(\phi)}{k^2} 
 \right)
\end{equation}
where  $'$   means derivative with respect to the field $\phi$. 
In order to determine the main properties of the flow
around the Gaussian FP, it is convenient to re-formulate
the equation in terms of dimensionless quantities, obtained by 
rescaling the 
dimensional variables in units of the scale $k$. 
Therefore we define ($k_0 $ is a boundary value for $k$)
\begin{equation}
t={\rm ln}\left (\frac{k_0}{k} \right ) \quad\quad\quad x=\frac{\phi}{k} 
\quad\quad\quad u(t,x) = \frac{U_k(\phi)}{k^4}
\label{dimensionless}
\end{equation}
and, by replacing them in Eq. \eqref{wh}, 
we get the dimensionless version of the Wegner-Houghton equation.
Then, FPs correspond to stationary (i.e. RG time $t$-independent)
solutions $u(x)^*$ of the dimensionless equation. In 
$d=4$ the only FP corresponds to the Gaussian solution $u^*=0$.
Finally, in order to determine the 
eigenvectors associated to the FP, we add to $u(x)^*$
a perturbation proportional to a small parameter $\delta$
and to the product of a generic function $w(x)$ and the exponential
factor  $e^{\lambda\, t}$ (with real $\lambda$):
\begin{equation}
u(x,t)= u(x)^* +  w(x)\,\delta \; e^{\lambda\, t} \; .
\label{eigendir}
\end{equation}
With the ansatz  in \eqref{eigendir} referred to the 
Gaussian FP  $u^*=0$, 
the dimensionless Wegner-Houghton equation at linear order 
in $\delta$, reads
\begin{equation}
\lambda w(x) +  x\, \frac{\partial  \, w(x) }{\partial x}
-4 w(x) =  \frac{\partial^2 \, w(x) }{\partial x^2\,}    
\label{lineq}
\end{equation}
which is a second order, homogeneous 
linear differential equation for 
$w(x)$, with  eigenvalue $\lambda$. 
By imposing the boundary 
$\partial_x w |_{x=0}=0$ associated to 
the symmetry requirement $w(x)=w(-x)$, 
and because of the redundancy of the overall 
normalization of  $w(x)$, due to the  
linearity of the equation, one could expect a 
unique solution for every value of $\lambda$.
Then, according to the specific form of 
$u(t,x)$ in \eqref{eigendir}, 
the sign of $\lambda$ determines whether 
the eigenmode is relevant ($\lambda>0$ and solution  
increasing  with $t$) or irrelevant 
($\lambda<0$ and solution decreasing with $t$) 
or marginal ($\lambda=0$ and no evolution with $t$). 

The direct resolution of Eq. \eqref{eigendir}, gives a
set of solutions with quantised eigenvalues $\lambda$
and another set with continuously varying $\lambda$. They 
can be summarized by (we shall make use of the notation adopted in
\cite{Branchina:2000}, and $c_0$ is an arbitrary normalization)
\begin{equation}
    w(x)=c_0 \left [ 1+ \sum_{n=1}^\infty
\frac{\Pi_{j=1}^n (\lambda +2 j-6) }{(2n)!}\; x^{2n}     
\right ]
\label{solgen}
\end{equation}
The quantised spectrum corresponds to $\lambda_h=4-2h$
with $h=0,1,2,3,...$, that produces a truncation of the 
infinite sum in \eqref{solgen} and therefore a polynomial
solution $w_h(x)$. For each integer value of $h$, 
$w_h(x)$ can be expressed in terms of the 
orthogonal normalizable basis of  the 
generalised Laguerre polynomials.

On the other hand, any other real value 
of $\lambda$, different from the above integers,
gives a non-polynomial infinite sum which can be expressed 
in terms of the Kummer (confluent hypergeometric) function
\begin{equation}
    M\left(\frac{\lambda-4}{2},\, \frac{1}{2},\,
    \frac{x^2}{2} \right ) \;.
    \label{kummer}
    \end{equation}

The behavior of \eqref{kummer} with $\lambda$ not 
belonging to the quantised spectrum,
is exponential at large values of the field 
$M(a,b,z)\sim {\rm e}^z\,/z^{(5-\lambda)/2}$,
differently from the polynomial solutions which  
grow as a power of $z$. Furthermore,
as remarked in \cite{Morris:2016}, due to the 
exponential behavior 
of the non-quantised of solutions in  \eqref{kummer}, 
these cannot be obtained from an expansion
in terms of the the generalised Laguerre polynomials.

As far as quantised polynomial solutions are concerned, 
it is easy to realize that for $h=0$ and $h=1$, one 
selects respectively a field independent  or a 
quadratic solution in the fields which, as  well known,
are relevant solutions. Then $h=2$ is a marginal 
(at tree level) solution quartic in the field that,
due to perturbative corrections, turns out to be irrelevant.
Finally all larger values of $h$ represent polynomial 
irrelevant solutions. 

The case of non-quantised solutions is more subtle; 
in fact, if one selects $\lambda >0$ 
(with $\lambda \neq 2$ and $\lambda \neq 4$), the
associated Kummer function in \eqref{kummer} is an 
independent solution (not representable as 
a  linear  combination of the quantised polynomials 
\cite{Bonanno:2000}) of 
the linear  equation \eqref{lineq}, with positive 
eigenvalue; therefore it corresponds, in principle, 
to a relevant eigenmode which allows to take the 
continuum  limit of the corresponding scalar theory,
as stated in \cite{HH2}. However, 
this conclusion is  false as first discussed in \cite{MorrisReb}
and thoroughly explained later in \cite{Morris:2016};
in fact, the exponential large field behavior of the
non-quantised solutions  makes the linear 
approximation adopted in Eq. \eqref{lineq} 
inadequate at large $x^2$ and consequently  
the $t$ evolution of these solutions derived from 
\eqref{lineq} is in contrast with 
the one derived from the complete flow equation. 
Actually, these solutions, at least for large values of
$x^2$, diverge in the UV limit $t\to -\infty$,
instead of vanishing as predicted by Eq. 
\eqref{lineq}, \cite{Morris:2016}.
Conversely, the solutions of the quantised spectrum have 
a  much smoother behavior at large $x^2$, and this implies that
 both the linear  approximation and  the full flow equation,
predict  the same $t$ evolution.

Then, statements on the $t$  evolution (and 
therefore on the relevance or irrelevance) deduced 
from Eq. \eqref{lineq}  concerning the 
quantised spectrum, are reliable as they are 
confirmed at the level of the full flow equation.
On the other hand,  the $t$ evolution of the 
solutions with non-quantised spectrum 
cannot be trusted as it is 
drastically modified when 
non-linear effects are taken into account \cite{Morris:2016}.

The solution \eqref{solgen} of  
the linearized flow equation \eqref{lineq} strongly
resembles the potential in \eqref{potential} and 
it is natural to apply the perturbative analysis 
developed in Sec. \ref{sec1}, to these eigenmodes. 
To this purpose we first need to put back the proper 
dimensions to the various parameters in \eqref{solgen}
and the dimensional potential corresponding to 
the dimensionless solution in Eq. \eqref{eigendir} 
(for the gaussian FP solution $u^*=0$), according to
Eq. \eqref{dimensionless}, is
\begin{eqnarray}
U_k(\phi) &=& \delta \; c_0 \,k^4 
\left (\frac{k}{k_0}\right )^{-\lambda}
\left [ 1+ \sum_{n=1}^\infty
\frac{\Pi_{j=1}^n (\lambda +2 j-6) }{(2n)!}\; 
\left (\frac{\phi}{k} \right )^{2n}\right ]
\nonumber\\
&= & k_0^4 \; h_0(k) \,
\left [ 1+ \sum_{n=1}^\infty
\frac{    h_{2n}(k) }{(2n)!}\; 
\left (\frac{\phi}{k_0} \right )^{2n}\right ]
\label{soldimen}  
\end{eqnarray}
where we introduced the couplings
\begin{equation}
h_0(k)=\delta \; c_0 \;
\left (\frac{k}{k_0}\right )^{4-\lambda}
\label{coupl1}
\end{equation}
\begin{equation}
h_{2n}(k)= \left (\frac{k_0}{k} \right )^{2n}
\; \Pi_{j=1}^n (\lambda +2 j-6) \; .
\label{coupl2}
\end{equation}

By identifying  $k_0$ with $M$ in \eqref{potential},
we observe that $U_k(\phi) $ in \eqref{soldimen} and 
$V(\phi)$ in \eqref{potential}  actually have similar structure.
However, there are also crucial differences;
beside an additional field independent 
term in \eqref{soldimen} which is subtracted
away in \eqref{potential},  the couplings 
in the two expressions have unlike arrangements.

In fact, while in $V(\phi)$  increasing powers of the 
same  coupling appear in front of the various vertices, 
$U_k(\phi)$   in Eq. \eqref{soldimen}  is globally
proportional to the only coupling $h_0(k)$ that contains 
undetermined parameters (the product $\delta \, c_0 $ and the
eigenvalue $\lambda$) and the couplings 
$h_{2n}(k)$, associated to every single vertex, 
have a well defined structure with no adjustable parameter. 
In addition, the 
dependence of $h_0(k)$ and $h_{2n}(k)$ on the scale $k$ is 
totally determined by the resolution of the linear flow equation. 
This latter property is essential in the study of the UV
properties of  $U_k(\phi)$.

If we first focus on the $k$-dependence of the couplings 
$h_{2n}(k)$ in Eq. \eqref{coupl2}, we notice that the powers of 
$1/k$  precisely compensate the powers of $\phi$ 
and the origin of this is the nature of the general 
solution in Eq. \eqref{eigendir}, written as the product  of two 
functions, each depending on one of the two dimensionless 
variables $t$ and $x$.
It is evident that, in the computation of the Green functions, 
the powers of $1/k$ in the couplings can compensate 
the divergence of  the tadpoles  generated by each 
vertex, and the sum of the tadpole series in each Green function
can be computed
(note however that  it does not sum up to an exponential as  
e.g. in \eqref{sumtadpole}, because of the numerical coefficients 
in the series).

Therefore for the  model  \eqref{soldimen} with $\lambda$ real but
not integer (otherwise the potential reduces to a polynomial 
whose renormalization properties are well known),
the sum of the tadpoles contributes to the vacuum fluctuations,
$2$-, $4$-, $6$-point Green functions, in  the following way:
\begin{align}
\label{tad1}
	{\cal V}_0 \qquad \; + \vcenter{\hbox{\includegraphics{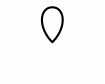}} 
      \vspace*{-1em}} + 
     \eqVcenter{\includegraphics{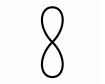}} + \; \dots \; &= 
     k^{4-\lambda}\,Q_0(k_0) \\ \label{tad2}
     \eqVcenter{\includegraphics{2p_0t_crossed}} + 
     \eqVcenter{\includegraphics{2p_1t}} + 
     \eqVcenter{\includegraphics{2p_2t}} + \; \dots \; &= 
    k^{4-\lambda}\,\frac{Q_2(k_0)}{k^2} \\  \label{tad3}	
    \eqVcenter{\includegraphics{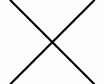}} + \eqVcenter{\includegraphics{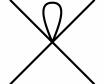}} + 
  \eqVcenter{\includegraphics{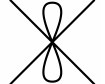}} + \; \dots \; &= 
   k^{4-\lambda}\,\frac{Q_4(k_0)}{k^4}   \\ \label{tad4}	
   \eqVcenter{\includegraphics{6p_0t}} + \eqVcenter{\includegraphics{6p_1t}} + 
     \eqVcenter{\includegraphics{6p_2t}} + \; \dots \; &= 
    k^{4-\lambda}\,\frac{Q_6(k_0)}{k^6}  
\end{align}
where in the first line ${\cal V}_0=k_0^4 \, h_0(k)$ 
is the zero-th order vacuum contribution from \eqref{soldimen}, 
and, in all equations, we factored out the $k$-dependent part of 
the sum and identified the UV cut-off $\Lambda$ of the tadpoles 
with the running scale $k$. 
The common factor $ k^{4-\lambda}$ comes from the overall
coefficient $h_0(k)$ in Eq. \eqref{soldimen}
and, finally,
$Q_0,\, Q_2, \, Q_4,\, Q_6$ are the $k$-independent factors 
resulting from the sum of the tadpoles. 

Whether Eqs. \eqref{tad1} -\eqref{tad4} vanish 
or diverge in the large $k$ limit clearly depends 
on  the eigenvalue $\lambda$. 
In addition,  we notice that Eqs. \eqref{tad1} -
\eqref{tad4} are indeed representative of the UV behavior 
of the respective $2n$-point Green functions because, 
as discussed at length in  Sec. \ref{sec1}, 
all other diagrams  containing more than one vertex 
are  subdominant  with  respect to the tadpole series, because of the 
suppression factor $O(1/k^{E+2V-4})$ in Eq. 
\eqref{eq:trend_generic_diagram}.

Since integer  values of $\lambda$, as discussed,
correspond to truncated polynomial potentials, whose 
renormalization properties are well known and do not contain 
asymptotically free eigenmodes, 
in the analysis of  Eqs. \eqref{tad1} -\eqref{tad4} 
we focus on non-integer $\lambda$.
We start with $4<\lambda$ that, in the limit $k \to \infty$,
clearly drives to zero Eqs.\eqref{tad1} -\eqref{tad4} 
and all higher $2n$-point functions,  with 
increasing powers of $1/k$.
If $2<\lambda<4$, again all  Green functions are vanishing,
with the exception of the sum of 
the vacuum diagrams that instead diverges; no further 
$k$-dependent parameter (or, in other words,
no further counterterm) is available to cure this divergence.
If $0<\lambda<2$, even the  $2$-point function becomes divergent
without any possible cure. Further lowering of $\lambda$ at 
negative values has the effect of making divergent more and more Green 
functions.

Then, if we look at the case $\lambda>0$ which,
according to \cite{HHOrig,HH2}, corresponds to asymptotically 
free modes, we find that there is at least 
one divergent, non-renormalizable, Green function if $\lambda<4$
(or, by subtracting away from the beginning the vacuum diagrams, 
the   non-renormalizable vertex appears for $\lambda<2$). 
Only when $\lambda>4$ (or $\lambda>2$ if the vacuum contribution 
is absent) divergent vertices are avoided,
but a suppressing factor in the  limit $k\to \infty$,
is present in all $2n$-point functions. In the next Section 
we show how this 
result can be compared  to the model in \eqref{potential}.

\section{RG approach to the Non-Polynomial toy model}
\label{sec3}

Before considering the full RG analysis of the model in 
Eq. \eqref{potential}, we can easily  determine the scale 
dependence of the only parameter
which gets renormalized in the scheme developed in Sec. \ref{sec1}, 
namely the coupling $g_R(\mu)$. In fact, 
from Eqs. \eqref{g0_renseries} and \eqref{delta1}
\begin{equation}
	g_0 = g_R + g_R^2 \, \frac{c}{M^2} \left( \mu^2 - \Lambda^2 - m^2 
 \log \frac{\mu^2}{\Lambda^2} \right) + O(g_R^3)
\end{equation}
where $c=1/(32\pi^2)$. Therefore,  since $g_0$ does not depend on 
the renormalization  scale $\mu$, the lowest order  $\beta$-function of 
the coupling  $g_R$ is
\begin{equation}
\label{betaper}
	\beta_g = \mu \, \frac{\partial g_R }{\partial \mu } 
 = - g_R^2 \, \frac{c}{M^2} \left( 2 \mu^2 - 2 m^2 \right)
\end{equation}
where the last term proportional to $m^2$ is  
neglected under the assumption $m^2<<\mu^2$.

Analogous result can be obtained from the full 
renormalization 
condition~\eqref{eq:renormalization}, since its left 
hand side is again $\mu$-independent, and therefore, 
by neglecting the logarithmic contribution we get
\begin{equation}
\label{betaper1}
	\beta_g = - \frac{2c \mu^2 g_R^2}{M^2 + g_R c \mu^2}
\end{equation}
which, in the lowest order  approximation, reproduces 
Eq. \eqref{betaper}.
Moreover, these $\beta$-functions reproduce as well 
the dependence of the bare coupling $g_0$ on the cut-off $\Lambda$.
We remark  that the sign of the  $\beta$-function in both 
\eqref{betaper} and  \eqref{betaper1} is negative.

Incidentally, we mention that the analysis  just described has been 
reconsidered also by using  dimensional regularization; in this case 
however we could not reach the same conclusions because
the $\beta$-function within this  scheme  turns out to be different 
from  \eqref{betaper1} and we also find inconsistencies as, 
for instance, the corresponding Lambert equation cannot be solved  
for positive $\epsilon =4-d$. 
This is not unexpected as our model contains an 
infinite number of dimensional couplings
that can be a source of problems if treated in dimensional 
regularization, \cite{branchina2}. Therefore, 
we prefer to retain the regularization procedure based on
the use of a UV cut-off that should grasp more accurately
the  physical content of this specific  problem.

Now we turn to a complete RG analysis of the model 
in \eqref{potential}, by means of the full (not linearized) 
Wegner-Houghton equation introduced in \eqref{wh}, that
in this case reads
\begin{equation}
	\label{eq:wh_our_potential}
 \frac{M^3 \phi\, k\dt g(k)}{2 \sqrt{g(k)}} \,
 \;\sinh \left( \sqrt{g(k)} \frac{\phi }{M} \right) 
 \; 	 = -  2 c k^4 \log 
 \left[ 1 + 
  \frac{g(k) \, M^2 }{k^2} \; 
 \cosh \left( \frac{\phi \sqrt{g(k)}}{M} \right) 
\right]
\end{equation}
where we introduced the notation $\dot g(k) = {d{g}(k)}/ {dk}$.

Eq. \eqref{eq:wh_our_potential} cannot be solved exactly for $g(k)$.
However, by assuming $g(k) <<1$,
an approximate solution can be obtained by 
selecting  the 
coefficients  of the various powers of the field $\phi$
on both sides of the equation
or, in other words, by considering the projections
of Eq. \eqref{eq:wh_our_potential} for the various 
vertices at zero external  momenta.
In particular, the flow of the 2-point 
function at zero momentum is obtained  by taking two 
derivatives with respect to $\phi$ and setting $\phi=0$. 
We get:
\begin{equation}
	\label{eq:beta_rg}
	k \dot g (k) = - \frac{2 c k^4 g^2(k)}{k^2 M^2 + M^4 
   g(k)}
\end{equation}
The resolution of the differential equation to the lowest order  gives:
\begin{equation}
\label{bet2}
	{g(k) = \frac{g(k_0)}{1 + \frac{c}{M^2} \, 
 g(k_0) \left( k^2 - k_0^2 \right)}}
\end{equation}
where $g(k_0)$ is the value of the coupling constant at the energy 
scale $k_0$  and, for large $k$, we get
$\dt{g} \sim - (k g^2)/M^2$  and
\begin{equation}
\label{trendg}
g \sim \frac{M^2}{k^2}
\end{equation}

Now, we are able to compare the running coupling found in the 
perturbative  and in the RG approach, namely in Eq. \eqref{betaper1}
and  in Eq. \eqref{eq:beta_rg}. 
We notice that, despite their agreement at 
lowest order in $g$, they differ at higher orders.
If we integrate Eq. \eqref{betaper1}, which was obtained  
from Eqs. \eqref{eq:renormalization},  we recover the
Lambert  function that, as already shown in Eq. \eqref{g_0asy},
for large values of energy goes like (the scale 
$\mu$ is here replaced by $k$)
\begin{equation}
\label{trendg2}
	g_R \sim \frac{\log k^2 - \log \log k^2}{k^2}
\end{equation}
Conversely, in the RG
approach we only get $g \sim 1/k^2$; the logarithmic corrections
of the perturbative series are missing and this mismatch is a 
consequence of the potential in Eq. \eqref{potential} being 
an approximate solution of the RG equation, so that 
the two approaches are compatible only up to these logarithmic terms.
Hence, the potential  \eqref{potential} 
does not identically satisfy the RG flow equation, but at least 
is a good approximate solution for large values of energy scale
(or equivalently for small values of the coupling constant).

Next, we analyze the flow after taking four derivatives  
with respect to the field in \eqref{eq:wh_our_potential} and,  
to the lowest-order in $g$, we get 
\begin{equation}
\label{bet4}
	g\,\dt{g} = - \frac{ c k g^3}{M^2} 
\end{equation}
which, evidently, does not produce the same differential flow equation
for $g$ as the one given in Eq. \eqref{eq:beta_rg}.
However,
if we regard  the left hand side of \eqref{eq:beta_rg} and
\eqref{bet4} as the differential flow of the $2$- and $4$-point 
Green function respectively (apart from some $k$-independent constants),
we see that  the 
the former is  of order $O(g^2)$ from  \eqref{eq:beta_rg}, and
the latter is $O(g^3)$ from \eqref{bet4}.
In other words, at large $k$ ($k^2>>M^2$),   
the discrepancy between 
\eqref{bet4} and  \eqref{eq:beta_rg} can be regarded as 
a higher order effect in $g$.

Green functions with larger number of external legs produce flow 
equations that are further suppressed. In fact, for
the $2n$-point Green functions at zero external 
momenta:
\begin{equation}
n\,g^{n-1}	\dt{g} = - \frac{2 c k g^{n+1}}{M^2}
\end{equation}
and  the suppression factor in the right hand side 
increases with $n$. According to the scaling
in \eqref{trendg}, the  $2n$-point  functions 
at large $k$ goes like $\sim (M^{2n}/k^{2n})$.

This result is explanatory in understanding the relation
between the potential in \eqref{potential} and the eigenmodes
of the linearized flow equation. In fact, if we use these
$k$-dependent  $2n$-point vertices to compute the renormalized sum 
of the tadpole series displayed in Eqs. \eqref{tad2} - \eqref{tad4}
(Eq. \eqref{tad1} is not considered here because in potential 
\eqref{potential} the vacuum  energy contribution is cancelled out),
we recover the result found in Sec. \ref{sec2}
for the specific value $\lambda=2$
(although it must be recalled   
that Eqs. \eqref{tad1} - \eqref{tad4}  in Sec. \ref{sec2}
require non-integer $\lambda$).
In other words, the potential in \eqref{potential}
represents a sort of peculiar structure  that reproduces the 
results in  Eqs. \eqref{tad2} - \eqref{tad4}  for
$\lambda=2$, which instead  is not covered by the the eigenmodes
of the linearized flow equation.
Then, we remark that in Eq. \eqref{tad2} with $\lambda=2$, 
any dependence on $k$ is cancelled and we are left with a 
finite renormalized $2$-point function, while  Eqs. 
\eqref{tad3}, \eqref{tad4}, \dots
depend on inverse powers of $k$ so that they vanish for $k\to\infty$,
which corresponds to the findings of  Sec. \ref{sec1}, i.e. a 
renormalized trivial theory with finite mass and vanishing 
interactions. On the other hand, the same computation with 
non-integer $\lambda>2$ yields an unphysical vanishing limit for all
relations  in \eqref{tad2} - \eqref{tad4} and higher order $2n$-point
vertices, i.e.  the model with all 
null renormalized Green functions is meaningless.


Before concluding this paragraph we comment on the 
physical implications associated to our findings.
By taking the point of view of  treating the model in \eqref{potential}
as an effective theory with a large
but finite  physical cut-off $k$, instead of ending with 
a trivial free theory,  we can write down an
effective potential at the scale $k$
\begin{equation}
\label{effpot}
	V(\phi) = \frac{1}{2} m_e(k)^2 \phi^2 + \frac{\lambda_e(k)}{4} 
      \phi^4
\end{equation}
where  higher powers of the field 
are suppressed by larger inverse powers of the scale $k$ 
and the effective mass and coupling can be related to 
the parameters of the original potential in \eqref{potential} 
through the  RG derived relation in \eqref{trendg}, 
\begin{equation}
\label{rel}
    m_e(k)^2=\frac{M^4}{k^2} \; , \quad\quad \quad
    \lambda_e(k)=\frac{M^4}{6\,k^4 }
    \end{equation}
By eliminating   $k$ in \eqref{rel}, we find
\begin{equation}
	\label{eq:M_of_mu}
	M = \frac{m_e(k)}{\sqrt[4]{6 \lambda_e(k)}}
\end{equation}
which generates  an effective  mass 
suppression mechanism: 
because of the interaction effects at large $k$, we expect a small
coupling $\lambda_e(k)$ and consequently a large ratio 
between the fixed mass  and the renormalized  effective mass,
$M/m_e(k)$. Moreover, the decreasing trend of $\lambda_e(k)$ with $k$ in
\eqref{rel}, marks a neat difference, potentially testable,
with respect to the scale dependence of the quartic coupling  
of the standard renormalizable $\phi^4$ theory.

\section{Conclusions}
\label{Conclusions}
We analysed a non-polynomial class of scalar potentials 
that allow to sum the series of tadpole diagrams, 
as this series contains the most severe UV  divergences, and  
taking care of the latter does guarantee the vanishing of the  
subleading divergences coming from multiple vertex diagrams.
In order to systematically classify
the diagrams according to their number of vertices, we adopted a
particular toy model that depends on one single  renormalizable 
coupling $g_0$  in such a way  that the potential is  an 
expansion in powers  of $g_0 \phi^2/M^2 $,
where $M$ is a fixed mass scale. This naturally allows  
for a perturbative treatment of the  radiative corrections.

Despite the   infinite tower of non-renormalizable couplings in  
\eqref{potential} the perturbative analysis shows that the model 
is renormalizable, but unfortunately, it is substantially trivial, 
as only  the 2-point Green function at zero  external momentum is 
finite, while all other vertices with at least four
external legs vanish in the  infinite UV cut-off limit 
$\Lambda \to \infty$. 
Although the effect of the interaction modifies the value of the  
mass scale $M$ into the renormalized  mass $m$, the model is 
practically indistinguishable from a free theory.
Slightly different non-polynomial interactions are also investigated
(see Appendix \ref{appB}), but  either they are  
non-renormalizable or, again, reproduce a trivial
renormalized theory.

The RG analysis shows a negative $\beta$-function of 
$g(k)$, produced by the  scaling  of  the  $2$-point function,
and we find  $g(k)\propto k^{-2}$, rather than 
$\propto 1/{\rm log} (k)$ as e.g.
for the  non-Abelian  Yang-Mills theory.  
Then, once the flow of $g(k)$ is established, 
the $2n$-vertices with $n>1$ proportional to 
$g^n$, scale  accordingly, i.e.  the larger $n$ 
the stronger the power-like suppression at large $k$.
It must be remarked that the model in \eqref{potential}
is not a full solution of the RG flow, as the numerical coefficient 
in front of the $g^n$ in each vertex with $n>1$
(the case $n=1$ is used to  determine the flow of $g$), 
is not consistent with the flow equation of that particular vertex.
However, due to the increasing power of $g$ with the power of the 
field in the vertices, one can regard the model
as an approximate solution with an error that is suppressed both by 
the absolute size of $g<<1$ and by the power $n$ of $g$, 
characteristic of each vertex.

If, according to the decreasing size of the
vertices for increasing $n$, we consider a two-parameter
effective model, valid  up to some UV scale $k$,
where all vertices with $n>2$ are neglected, 
we find a very weakly interacting theory.
However, unlike the  two-parameter renormalizable standard $\phi^4$
scalar theory, in this case the effective coupling 
decreases in the UV limit and, potentially, 
this feature has  experimentally testable consequences.

The potentials  introduced in \cite{HHOrig,HH2}, can 
be included in our analysis due to the similarities in the form 
of the non-polynomial interaction
which allows to sum the tadpole diagrams. 
However, the scale dependence of these potentials is entirely
established by the resolution of the flow equation, 
linearized around the Gaussian fixed point.
Once  these eigenmodes are converted  in  dimensional form, 
one finds that each  vertex is normalized by  the proper 
inverse power of the running scale $k$, in addition to 
an overall factor $k^{4-\lambda}$, exhibiting the eigenvalue
$\lambda$ which establishes the UV properties of the solution.

These solutions cannot be retained as asymptotically 
free  eigenmodes, because they do not respect the 
linear  approximation at large values of the field and 
consequently the predicted  RG-time evolution is wrong, 
 as clearly explained in \cite{Morris:2016}.
The counterpart of these conclusions within our perturbative 
analysis is that these eigenmodes, depending on 
the value of $\lambda$, are either not 
renormalizable, because of some residual divergence,
or unphysical, because all Green functions are vanishing.
Furthermore, these eigenmodes do not include the case of 
integer $\lambda$, while we found that the  potential
in \eqref{potential} corresponds to the scaling
of Eqs. \eqref{tad1} - \eqref{tad4} with $\lambda=2$.

\section*{Acknowledgements}

The authors benefited from fruitful discussions with A. Bonanno, 
V. Branchina, F. Contino.
This work has been carried out within the INFN project FLAG.

\appendix

\section{}
\label{appA}
In this appendix  we analyze the leading  divergence 
of the sum of the melonic-like diagrams $\hat S$, depicted in
Eq. \eqref{maggior}, in cut-off  regularization.
First, let us define:
\begin{equation}
	\Delta_k = \frac{1}{k^2+m^2}
\end{equation}
Thus, the first integral of the series $\hat S$, i.e. the
sunset diagram with external momentum $p$ and loop 
momenta $a$ and $b$, is given by:
\begin{equation}
\label{sunset}
	\begin{gathered}
		S(p) \equiv \eqVcenter{\includegraphics{2p_2m}} = \frac{g_0^4}{3!} \int 
  \frac{d^4a}
  {(2 \pi)^4} \int \frac{d^4b}{(2 \pi)^4} \, \Delta_a \, \Delta_b \, 
  \Delta_{a+b+p} 
  = \\
		= \frac{g_0^4}{6 (4 \pi)^2} \left[ 3 \Lambda^2 + p^2 
      \left( \alpha 
   + \beta 
  \log \frac{\Lambda^2}{m^2} \right) + O \left(\log 
  \frac{\Lambda^2}{m^2}\right) + 
  Q(p,m) \right]
	\end{gathered}
\end{equation}
where the result of the integration is taken from~\cite{IIM}, 
and $\alpha$ and $\beta$ are constants. 
We will neglect the contribution  $O \left(\log \frac{\Lambda^2}
{m^2}\right)$ because it is clearly subdominant if compared to $\Lambda^2$. 
In addition, important contributions could come from $Q(p,m)$ 
when we use the above 
result for the subsequent diagrams (see below); however it is 
straightforward 
to realize that, when $p \sim \Lambda$, the most divergent contribution 
is proportional to $\Lambda^2$:
\begin{equation}
	\int \frac{d^4a}{(2 \pi)^4} \int \frac{d^4b}{(2 \pi)^4} \, \Delta_a \, \Delta_b 
 \, \Delta_{a+b+p} \quad \overset{p \sim \Lambda}{\longrightarrow} \quad \frac{1}
 {\Lambda^2} \int \frac{d^4a}{(2 \pi)^4} \int \frac{d^4b}{(2 \pi)^4} \, \frac{1}{a^2 
 b^2} \sim \Lambda^2
\end{equation}
In other words, there is no contribution that goes like  $\Lambda^2 \log 
\frac{\Lambda^2}{m^2}$ and, in the following, we can  neglect  terms proportional to $\beta$
in \eqref{sunset}.

Let us now compute the subsequent diagram of the series, 
this time with external momentum $p=0$ and with $e$ and $f$ as the
additional loop momenta:
\begin{equation}
\label{bitadpole}
	\begin{gathered}
		\eqVcenter{\includegraphics{2p_4m}} = \frac{g_0^6}{5!\, M^4 } \int \frac{d^4a}{(2 \pi)^4} 
  \int \frac{d^4b}{(2 \pi)^4} \int \frac{d^4e}{(2 \pi)^4} \int \frac{d^4f}{(2 
  \pi)^4} \, \Delta_a \, \Delta_b \, \Delta_e \, \Delta_f \, \Delta_{a+b+e+f} = \\
   = \frac{g_0^6}{5!\, M^4 } \int \frac{d^4e}{(2 \pi)^4} \int \frac{d^4f}{(2 \pi)^4} \left[ 
   3! \, S(e+f) \right] \Delta_e \, \Delta_f = \\
   = \frac{g_0^6}{5!\, M^4 } \int \frac{d^4e}{(2 \pi)^4} \int \frac{d^4f}{(2 \pi)^4} \, 
   \Delta_e \, \Delta_f \left[ \frac{3 \Lambda^2}{(4 \pi)^2} + \frac{\alpha}{(4 
    \pi)^2} \left(e+f\right)^2 \right] = \\
    = \frac{g_0^6}{5!\, M^4 } \int \frac{d^4e}{(2 \pi)^4} \int 
    \frac{d^4f}{(2 \pi)^4} \, 
    \Delta_e \, \Delta_f \left[ c_1 I + c_2 \left( e^2 + f^2 + 2 e f \cos 
    \theta_{ef} \right) \right]
	\end{gathered}
\end{equation}
where $\theta_{ef}$ is the angle between momenta $e$ and $f$, $I$ is the tadpole 
integral already evaluated in  Eq. \eqref{eq:I}, and we  defined:
$c_1 = 6$ and  $c_2 = {\alpha}/{(4 \pi)^2}$.

By exploiting the symmetry in the last line in Eq. \eqref{bitadpole}
we find:
\begin{gather*}
\eqVcenter{\includegraphics{2p_4m}} = \frac{g_0^6}{5!\, M^4 } \int 
\frac{d^4e}{(2 \pi)^4} \int 
\frac{d^4f}{(2 \pi)^4} \, \Delta_e \, \Delta_f \left( c_1 I + 2 c_2 e^2 
\right) = \\
= \frac{g_0^6}{5!\, M^4 } \left( c_1 I^3  + 2 c_2 I
\int\frac{d^4e}{(2 \pi)^4} \, \frac{e^2}{e^2 + m^2} \right) = \\
	= \frac{g_0^6}{5!\, M^4 } \left( c_1 I^3 + 2 c_2 I\, c_3 \Lambda^4 
 \right) = \frac{g_0^6}{5!\, M^4 } \left( c_1 I^3 + 2 c_2 c_4 I^3 \right)
\end{gather*}
where $c_3$ and $c_4$ are constants. By following the same
procedure with the successive diagram of the series $\hat S$,  we get
\begin{equation}
	\eqVcenter{\includegraphics{2p_6m}} = \frac{g_0^8}{7! \, M^8} 
 \left( c_1 I^5 + 4 c_2  c_4 I^5 \right)
\end{equation}
and so on and so forth for all the other diagrams of the series.

Before adding everything up, we have to remember that each vertex 
must be "dressed" 
with the tadpole diagrams sum, and this means that a global factor $g_0^2$ (a 
factor $g_0$ for each vertex) must be discarded from the global count of the
powers of $\Lambda$ in each diagram of $\hat S$. Then, the sum reads
\begin{equation}
	\begin{gathered}
		\eqVcenter{\includegraphics{2p_2m}} + \eqVcenter{\includegraphics{2p_4m}} + 
  \eqVcenter{\includegraphics{2p_6m}} + \dots \sim \\
		\sim g_0^2 I + \frac{g_0^4}{5!\, M^4} \left( c_1 + 2 c_2 c_4 
  \right) 
  I^3 + \frac{g_0^6}{7!\, M^8} \left( c_1 + 4 c_2 c_4 \right) I^5 + \dots
   = \\
	g_0^2 I + c_1 \left( \frac{g_0^4}{5!\, M^4} I^3 + \frac{g_0^6}{7!\,M^8} 
 I^5 + \dots \right) + c_2 c_4 \left( \frac{2}{5!\,M^4} g_0^4 I^3 + 
 \frac{4}{7!\, M^8} g_0^6 I^5 + \dots \right) < \\
	< g_0^2 I + \frac{c_1,M^4}{I} \left( \frac{g_0^4 I^4}{4!\, M^8}
 +\frac{g_0^6 I^6}{6!\,M^{12}} + \dots \right) + \frac{c_2 c_4\,M^4}{I} 
 \left( 
 \frac{g_0^4 I^4}{4!\,M^8} + \frac{g_0^6 I^6}{6!\,M^{12}} + \dots \right) 
 = \\
		= g_0^2 I + \frac{b\,M^4}{I} \left[ \cosh \left( \frac{g_0 I }{M^2}
  \right) - 1 - \frac{g_0^2 I^2}{2\,M^2} \right]
	\end{gathered}
 \label{inequality}
\end{equation}
where in the last step we have defined $b = c_1 + c_2 c_4$.
The inequality derived in Eq. \eqref{inequality} 
is the one reported in Eq. \eqref{maggior}.


\section{}
\label{appB}

In principle, the potential  studied in Sec. \ref{sec1} can be modified 
to adjust its UV behavior with the aim of 
obtaining  a significant interacting theory.
Unfortunately, in all cases here analyzed, the resulting model  
turns out to be either not practicable or equivalent to
the original one and, below, we discuss a few
representative cases.

Actually, one of these models is already considered in Sec. \ref{sec2}. 
It is the one coming from  the direct resolution of the RG flow equation, 
suitably linearized around the Gaussian fixed point, that can be expressed 
as an expansion in powers of the field but with a rather different 
structure of  the couplings.

Another model,  alternative to \eqref{potential}, but very similar 
in structure, is given by
\begin{equation}
\label{potential1}
	V(\phi) = \frac{M^4}{g_0} \left( \frac{g_0 \phi^2}{2! M^2} + \frac{g_0^2 
 \phi^4}{4! M^4} + \frac{g_0^3 \phi^6}{6! M^6} + \dots \right) = \frac{M^4}
 {g_0} \left[ \cosh \left(\sqrt{g_0} \frac{\phi}{M}\right) - 1 \right]
\end{equation}
where, as in \eqref{potential}, we do not renormalize the parameter M 
(i.e. it contains no counterterms), and 
the only renormalizable parameter is $g_0$,
so that the only difference with respect to the original model is an overall
rescaling of the factor $1/g_0$. Then, by repeating the same analysis 
of Sec. \ref{sec1}, we immediately realize that the sum of the tadpoles
(the most divergent diagrams), in the two point function
yields $M^2 \, e^{g_0 I / M^2}$, which is  equal to the result 
in \eqref{2pttad}, up to the mentioned rescaling of $1/g_0$ and the same 
result holds for all $2n$-point  Green functions.
It is clear that the renormalization condition adopted in Sec. \ref{sec1}
in this case would be insufficient to make the $2$-point Green function 
finite and we need a different prescription. In particular, if  we take
\begin{equation}
\label{2ptalt1}
	M^2 e^{g_0 I / M^2} = M^2 e^{g_R c \mu^2 / M^2} \qquad \implies \qquad 
 e^{g_0 I / M^2} = e^{g_R c \mu^2 / M^2}
\end{equation}
we find the quite simple dependence of $g_0$ on the UV cut-off $\Lambda$
\begin{equation}
\label{newrenor}
	g_{0} (\Lambda) = g_R \, \frac{\mu^2}{\Lambda^2} 
\end{equation}
which is the same as the  one found in Eq. \eqref{g_0asy}, 
apart from the logarithmic corrections.
However, the result in \eqref{newrenor} means that it is not possible to 
frame this renormalization scheme within the standard series expansion 
of $g_0$ in powers of $g_R$; 
in other words, here we do not have the usual cancellation of the divergence 
through the subtraction of an equally divergent counterterm, 
and we are forced to adopt a multiplicative cancellation.

By following the same power counting analysis performed in Sec. \ref{sec1},
we find  that  the relation displayed in \eqref{newrenor},
leaves finite  the sum of the tadpole series in the  $2$-point Green function
and forces to zero all the subleading non-tadpole diagrams. 
In addition, the tadpole series contributing to all other $2n$-point 
functions identically vanishes, because of further 
suppressing multiplicative powers of $g_0$; 
as a consequence all  but the $2$-point Green function are null.
Then, we conclude that the renormalized model \eqref{potential1}
is identical to the renormalized \eqref{potential},
although the latter admits a perturbative treatment in terms of 
counterterms which cannot be applied to the former.

A different approach to the renormalization of this kind of models is 
instead obtained by allowing for the renormalization of the 
other parameter appearing in 
the potential, namely $M$. So, for instance, we can reconsider the 
previous 
case  but with the substitution $M \to M_0$, i.e. a bare mass that 
undergoes 
renormalization:
\begin{equation}
	V(\phi) = \frac{M_0^4}{g_0} \left( \frac{g_0 \phi^2}{2! M_0^2} + 
 \frac{g_0^2 \phi^4}{4! M_0^4} + \frac{g_0^3 \phi^6}{6! M_0^6} + \dots 
 \right) 
 = \frac{M_0^4}{g_0} \left[ \cosh \left(\sqrt{g_0} \frac{\phi}{M_0}\right) 
 - 1 
 \right]
\end{equation}
Of course, the sum of the tadpoles discussed in the previous case still 
holds, 
but now  we find  the ratio $g_0/M_0^2$ instead of $g_0/M^2$
in the  exponent. 

If we now look at the output of the tadpole sum in 
the $2$-point and $4$-point
functions we get $M_0^2 e^{g_0 I / M_0^2}$ and 
$g_0 e^{g_0 I / M_0^2}$ respectively. 
These results are both finite only if the same 
counterterms  are taken for $g_0$ and $M_0^2$.
Then, these counterterms get
cancelled in the ratio $g_0/ M_0^2$ but they can be chosen in such a way
that the tadpole series in the $2$-point  and $4$-point functions is 
finite.
In this case however,
we also find that the tadpole  series remains finite
in any $2n$-point function, as for instance in the $6$-point function
\begin{equation}
	\begin{gathered}
		\eqVcenter{\includegraphics{6p_0t}} + \eqVcenter{\includegraphics{6p_1t}} + 
  \eqVcenter{\includegraphics{6p_2t}} + \; \dots \; = \frac{g_0^2}{M_0^2} \, 
  e^{g_0 I / 
  M_0^2}
  = \underbrace{\frac{g_0}{M_0^2}}_{\text{finite}} \, \underbrace{g_0 \, 
  e^{g_0 I 
  / M_0^2}}_{\text{finite}} 
	\end{gathered}
\end{equation}
This, in turn, implies that any $2$-vertex diagram 
(like for instance the sunset
diagram) with the tadpole series summed at each vertex, 
produces a divergence 
that cannot be cured because all counterterms coming from $g_0$ and $M_0^2$
have already been used.

We notice that even considering the possibility of renormalizing the
mass term $M_0$ in the original potential in 
\eqref{potential}, i.e. 
\begin{equation}
	V(\phi) = M_0^4 \left( \frac{g_0 \phi^2}{2! M_0^2} + \frac{g_0^2 
  \phi^4}{4! M_0^4} + \frac{g_0^3 \phi^6}{6! M_0^6} + \dots 
 \right) = M_0^4 \left[ \cosh \left(\sqrt{g_0} \frac{\phi}{M_0}\right) - 1 
 \right]
\end{equation}
we would get  $g_0\, M_0^2 e^{g_0 I / M_0^2}$ and $g^2_0 e^{g_0 I / M_0^2}$
as  output of the tadpole sum in the $2$-point and $4$-point functions and
consequently, as before, the same renormalization counterterms for 
$g_0$ and $M^2_0$ are required.
Then,  the same flaw encountered in the previous example
shows up even in this case, namely the requirement to fix
the counterterms by imposing a finite tadpole series, 
implies  as well the presence of 
uncured divergences in the diagrams containing at least two vertices, 
such as the sunset diagram.


\end{document}